\documentclass[preprint,12pt]{elsarticle}

\usepackage{amssymb}
\usepackage{amsmath}
\usepackage{hyperref}
\usepackage[export]{adjustbox}
\usepackage{subcaption}
\usepackage{algorithm,algorithmicx,algcompatible}
\usepackage{algpseudocode}
\usepackage{amsmath,bm,xspace}
\usepackage{multirow}
\usepackage{booktabs}
\usepackage{amssymb} 
\usepackage{pifont}

\usepackage[final]{changes}

\newcommand{\emb}[1]{\bm{#1}}
\newcommand{\setOm}{\Omega_{\emb{e}_S}}
\newcommand{\id}{\textsf{id}\xspace}

\journal{Knowledge Base System}

\begin{document}

\begin{frontmatter}



\title{STCKGE: Continual Knowledge Graph Embedding Based on Spatial Transformation}


\author[1]{Xinyan Wang\fnref{label1}}
\ead{wangxinyan@whu.edu.cn}
\fntext[label1]{orcid: 0009-0000-7835-4296}

\author[1]{Jinshuo Liu\corref{cor1}}
\ead{liujinshuo@whu.edu.cn}
\cortext[cor1]{Corresponding Author}

\author[1]{Kaijian Xie}[]
\ead{kaijianXie@whu.edu.cn}

\author[1]{Meng Wang}[]
\ead{wang_meng@whu.edu.cn}

\author[1]{Cheng Bi}[]
\ead{bicheng@whu.edu.cn}

\author[1]{Juan Deng\corref{cor1}}[]
\ead{dengjuan@whu.edu.cn}

\author[1]{Donghong Ji}[]
\ead{dhji@whu.edu.cn}

\author[2]{Jeff Z. Pan}[]
\ead{j.z.pan@ed.ac.uk}

\affiliation[1]{
    organization={School of Cyber Science and Engineering, Wuhan University},
    addressline={299 Bayi road}, 
    city={Wuhan},
    postcode={430072}, 
    country={China}}

\affiliation[2]{
    organization={Knowledge Graph Group, Alan Turing Institute, The University of Edinburgh},
    addressline={57 George Square}, 
    city={Edinburgh},
    postcode={EH8 9JU}, 
    country={U.K}}

\begin{abstract}

Current Continual Knowledge Graph Embedding (CKGE) methods primarily rely on translation-based embedding approaches, leveraging previously acquired knowledge to initialize new facts. While these methods often integrate fine-tuning or continual learning strategies to enhance efficiency, they compromise prediction accuracy and lack support for complex relational structures (e.g., multi-hop relations). To address these limitations, we propose STCKGE, a novel CKGE framework \replaced{based on}{grounded in} spatial transformation. In this framework, entity positions are jointly determined by base position vectors and offset vectors\replaced{, enabling the model to represent complex relations more effectively while supporting efficient embedding updates for both new and existing knowledge through simple spatial operations, without relying on traditional continual learning techniques.}{. This design not only enhances the representation of complex relational structures but also enables efficient embedding updates for both new and old knowledge through simple spatial transformations, eliminating the need for traditional continual learning methods.} Furthermore, we introduce a bidirectional collaborative update strategy and a balanced embedding method to \replaced{guide}{refine} parameter updates, effectively minimizing training costs while improving model accuracy. We comprehensively evaluate our model on seven public datasets and a newly constructed dataset (MULTI) focusing on multi-hop relationships. Experimental results confirm STCKGE's strong performance in multi-hop relationship learning and prediction accuracy, with an average MRR improvement of 5.4\%. Our code and dataset are available at \href{stckge}{https://github.com/Wxy13131313131/STCKGE}
\end{abstract}


\begin{highlights}
\item Proposes STCKGE, a novel spatial transformation-based CKGE framework featuring dual-component entity representations (base vector + offset vector) and relation-aware spatial regions, significantly enhancing complex relational modeling (e.g., multi-hop) while reducing new knowledge dependency on historical embeddings.

\item Introduces a Bidirectional Collaborative Update (BCU) strategy that efficiently propagates knowledge through lightweight offset vector operations, minimizing retraining costs for both new and historical knowledge.

\item Experimental results confirm STCKGE's strong performance in multi-hop relationship learning and prediction accuracy, with an average MRR improvement of 5.4\%.

\item Constructs and releases the MULTI benchmark dataset with explicit multi-hop facts, addressing selection bias in existing CKGE benchmarks.

\end{highlights}


\end{frontmatter}



\section{Introduction}
\added{Although knowledge graphs (KGs) are highly effective for representing structured data, their symbolic nature often makes them difficult to manipulate\cite{DBLP:journals/tkdd/RossiBFMM21}.}
\replaced{The emergence of knowledge graph embedding (KGE) models marked a major step forward by enabling the representation of entities and relations in a low-dimensional vector space\cite{DBLP:conf/kdd/0001GHHLMSSZ14}.}{Knowledge Graph Embedding (KGE) models  represent a significant advancement in modeling entities and relationships from multi-relational data within low-dimensional vector spaces \cite{DBLP:conf/kdd/0001GHHLMSSZ14}.} 
These models are trained on known facts to learn embeddings for entities and relations. The learned embeddings facilitate the computation of plausibility scores for potential facts, thereby establishing a robust structural foundation for numerous knowledge-based applications. 
\deleted{Among early influential models, TransE pioneered a translational approach, evaluating fact plausibility by modeling relationships as translations between head and tail entity vectors.} 
\replaced{As research progressed, numerous KGE models emerged, each offering improvements from different perspectives.}{Subsequent research has yielded a diverse array of KGE models, offering various optimizations and extensions from distinct perspectives.}\cite{Gutierrez-Basulto18,TrouillonWRGB16,SunDNT19, WangMWG17, DBLP:conf/kcap/HubertPMBM23}
\replaced{However, Gutierrez-Basulto et al. \cite{Gutierrez-Basulto18} pointed out that even for the simplest types of relations, translation-based KGE models often fail to capture relational patterns effectively. This finding revealed key limitations of existing KGE approaches and spurred the development of new methods.}{However, studies such as \cite{Gutierrez-Basulto18} have demonstrated limitations in relational modeling capabilities for even simple relationship types within translation-based KGE models. This finding underscores the constraints of existing KGE approaches and has motivated new research directions.}

 To address these shortcomings, Space-Translation-Based KGE methods \cite{abboud2020boxe} have been introduced, enhancing relational modeling through more complex spatial transformations and relation representations. Despite these developments, the dynamic nature of knowledge growth presents a critical challenge for large-scale static KGE methods \cite{TransE,DBLP:conf/emnlp/PanW21,DBLP:conf/coling/LiuWLSX20}. In response, the Continual Knowledge Graph Embedding (CKGE) task has emerged, providing a framework for constructing and applying dynamic knowledge graphs.

\begin{figure}[!t]
	\centering
	\includegraphics[width=0.75\textwidth]{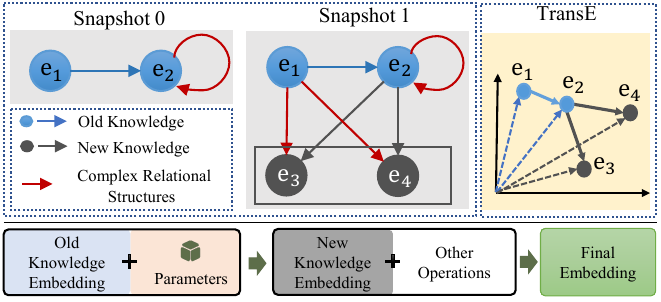} 
	\caption{Limitations of mainstream CKGE frameworks. The top section demonstrates how current methods, such as TransE, fail to represent complex relational structures. The bottom section illustrates the reliance of mainstream CKGE training frameworks on old knowledge for representing new information. }
	\label{fig:1}
\end{figure}

CKGE aims to model dynamic knowledge \replaced{ by integrating new entities and relationships while effectively preserving existing knowledge}{by accommodating knowledge growth through the integration of new entities and relationships while effectively preserving existing knowledge}\cite{kirkpatrick2017overcoming}. The \replaced{main}{core} challenge \replaced{is to assimilate existing knowledge efficiently, ensuring that new facts are accurately represented while maintaining alignment between historical and emerging knowledge.}{lies in efficiently assimilating existing knowledge to accurately represent novel facts while maintaining seamless alignment between historical and emerging knowledge representations.} \replaced{Current}{Prevailing} CKGE methods primarily use translation-based embedding techniques,\replaced{whitch are enhanced by} dynamic architectures, memory replay, and regularization mechanisms to incorporate new entities and facts on prior knowledge\cite{DBLP:conf/semweb/ShabanP24,LiuK0SGLJL24,LiuK0WGSL0JL24,CuiWSLJHH23}. Although these approaches enhance learning efficiency, they often reduce predictive accuracy. This limitation arises from their heavy reliance on historical knowledge when acquiring  new knowledge and \replaced{their inability to effectively model complex relational structures, such as multi-hop relationships,}{their inherent incapacity to model complex relational structures (e.g., multi-hop relationships),} as illustrated in Figure 1.

To strike a balance between learning efficiency and predictive accuracy while improving the representation of complex relational structures, we propose a new CKGE framework based on spatial transformations. Inspired by the embedding model in BoxE\cite{abboud2020boxe}, we define entity positions using both base position vectors and offset vectors. This dual-representation mechanism reduces the dependency of new knowledge on old knowledge, facilitates dynamic interactions between old and new knowledge, and enables the model to better capture logical rules and represent complex relational structures. For learning new knowledge, we introduce a bidirectional collaborative update strategy. By updating offset vectors with simple operations, we achieve efficient embedding updates for new and old knowledge\replaced{, minimizing the need for retraining. }{while minimizing retraining costs.} To tackle the long-standing issue of catastrophic forgetting, we also propose an innovative balancing embedding method. This approach  applies fine-grained constraints on parameter updates, thereby improving the model's predictive accuracy.

Additionally, existing CKGE datasets often exhibit selection bias toward facts directly linked to current knowledge, \replaced{which can lead to overfitting and hinder the rigorous evaluation of models' ability to handle multi-hop reasoning and novel knowledge}{potentially inducing overfitting and impeding rigorous evaluation of models' capabilities in handling multi-hop relationships and novel knowledge}. To support comprehensive and objective assessment, we construct a new benchmark dataset, MULTI, specifically designed to include entities involved in multi-hop relationships. \replaced{This allows for deeper evaluation of models' capabilities in knowledge assimilation and integration under continual learning settings.}{This facilitating in-depth investigation of models' knowledge assimilation and integration capabilities in continual learning scenarios.} Extensive experiments conducted on \replaced{seven}{four} public datasets and our MULTI benchmark demonstrate that the proposed model consistently outperforms state-of-the-art baselines.

The specific contributions of this work are as follows:
\begin{itemize}
    \item We propose a new region-based embedding framework, enable the model to capture complex relational structures more effectively. \replaced{To the best of our knowledge, this is the first application of space-translation-based methods to the CKGE task.}{This represents the first attempt to apply space-translation-based methods to the CKGE task.}
    \item \replaced{We design a bidirectional collaborative update strategy that updates embeddings for both new and existing knowledge via spatial offset operations, significantly reducing retraining costs.}{ We define a bidirectional collaborative update strategy, which updates the embeddings of old and new knowledge through spatial offset vector operations and reduces the retraining costs.} 
    \item We introduce a balancing embedding method to \replaced{mitigate}{address} catastrophic forgetting, improving predictive accuracy. 
    \item Extensive experimental results \replaced{show}{demonstrate} that our model enhances predictive accuracy while maintaining high learning efficiency. Notably, on the MULTI, STCKGE achieved significant improvements in MRR, Hits@1, Hits@3, and Hits@10, with increases of 5.7\%, 6.7\%, 6.7\%, and 7.9\%, respectively. Furthermore, we have released a newly constructed CKGE dataset MULTI that integrates complex relational structures.
\end{itemize}

\section{Related Work}

\textbf{Traditional KGE:} Traditional Knowledge Graph Embedding (KGE) models \replaced{are generally categorized into translation-based and text-based approaches.}{primarily fall into two categories: translation-based methods and text-based methods.} Translation-based methods map entities and relations into low-dimensional vector spaces. Pioneering work includes TransE \cite{TransE}, which formulates the tail entity as the summation of the head entity and relation vector. Subsequent extensions encompass rotation-based models such as RotatE \cite{SunDNT19} that defines relations through rotations in complex vector spaces, and ComplEx \cite{TrouillonWRGB16} operating in complex number spaces. Conversely, text-based methods \cite{DettmersMS018, DBLP:conf/naacl/NguyenNNP18} leverage semantic information for representation learning. MASCHInE\cite{DBLP:conf/kcap/HubertPMBM23} proposes heuristics for creating semantic-aware protographs to learn improved knowledge graph embeddings that better capture KG semantics. Nevertheless, these conventional methods exhibit limitations in modeling complex relational structures.

To address these challenges, hybrid spatial-translation methods have emerged \cite{Gutierrez-Basulto18}. Notable among these is BoxE\cite{abboud2020boxe}, which introduces explicit spatial regions to represent relations, thereby capturing intricate relational patterns. \added{BoxE demonstrates stronger support for complex logical relations, as shown in Table \ref{tab:pattern-support}.} However, such techniques remain confined to static KGE settings and cannot accommodate dynamic knowledge growth.

\begin{table}[ht]
\caption{Support for different logical relations by BoxE and TransE. \checkmark/\checkmark: fully supported, \checkmark/\ding{55} or \ding{55}/\checkmark: partially supported, \ding{55}/\ding{55}: unsupported.}
\label{tab:pattern-support}
\centering
\renewcommand{\arraystretch}{0.8}
\setlength{\tabcolsep}{10pt}
\adjustbox{width=0.6\textwidth}{
\begin{tabular}{l|c|c}
\toprule
\textbf{Logical Relations} & \textbf{BoxE} & \textbf{TransE} \\
\midrule
Symmetry: $r_1(x, y) \Rightarrow r_1(y, x)$ & \checkmark/\checkmark & \ding{55}/\ding{55} \\
Anti-symmetry: $r_1(x, y) \Rightarrow \lnot r_1(y, x)$ & \checkmark/\checkmark & \checkmark/\checkmark \\
Inversion: $r_1(x, y) \Leftrightarrow r_2(y, x)$ & \checkmark/\checkmark & \checkmark/\ding{55} \\
Composition: $r_1(x, y) \land r_2(y, z) \Rightarrow r_3(x, z)$ & \ding{55}/\ding{55} & \checkmark/\ding{55} \\
Hierarchy: $r_1(x, y) \Rightarrow r_2(x, y)$ & \checkmark/\checkmark & \ding{55}/\ding{55} \\
Intersection: $r_1(x, y) \land r_2(x, y) \Rightarrow r_3(x, y)$ & \checkmark/\checkmark & \checkmark/\ding{55} \\
Mutual exclusion: $r_1(x, y) \land r_2(x, y) \Rightarrow \bot$ & \checkmark/\checkmark & \checkmark/\checkmark \\
\bottomrule
\end{tabular}}

\end{table}


\textbf{CKGE:}
Mainstream CKGE methods \cite{kirkpatrick2017overcoming,rusu2016progressive, lomonaco2017core50,lopez2017gradient,DBLP:conf/naacl/WangXYGCW19,zenke2017continual} typically adopt TransE as the \replaced{foundational }{core} embedding framework, augmented with fine-tuning or continual learning techniques. Specific implementations include disentangled architectures with dual modules for knowledge updates \cite{KouLLLZZ20}, hypernetwork-injected parameters for adaptive modeling \cite{DBLP:conf/aaai/00080TCLC24}, GCN-based autoencoders with embedding transfer strategies \cite{CuiWSLJHH23}, and incremental distillation mechanisms with lightweight adapters \cite{LiuK0WGSL0JL24, LiuK0SGLJL24}. Despite improving new snapshot integration efficiency, these methods significantly compromise predictive accuracy. More critically, they fundamentally struggle to assimilate new knowledge lacking direct associations with existing structures. Consequently, developing balanced embedding strategies that simultaneously ensure efficiency and accuracy remains an urgent challenge in CKGE research
\section{Methodology}
\subsection{Preliminary }
\paragraph{Growing Knowledge} Dynamic knowledge graphs are represented as a sequence of evolving snapshots, i.e., $\mathcal{G} ={\mathcal{S}_1,\mathcal{S} _2,\dots,\mathcal{S}_n}$. Each snapshot $\mathcal{S}_i$ is represented as a triplet $(\mathcal{E} _i, \mathcal{R}_i, \mathcal{T}_i)$, where $\mathcal{E} , \mathcal{R}$ and $\mathcal{T}$ denote the sets of entities, relations, and facts at time $i$, respectively. $\Delta \mathcal{E}_i = \mathcal{E}_i-\mathcal{E}_{i-1}$, $\Delta \mathcal{R}_i = \mathcal{R}_i-\mathcal{R}_{i-1}$ and $\Delta \mathcal{T}_i = \mathcal{T}_i-\mathcal{T}_{i-1}$ are denoted as new entities, relations, and facts.

\paragraph{Continual Knowledge Graph Embedding} CKGE aims to map newly added entities and relations into the vector space while ensuring a balanced update with old entities and relations. In more details, for each snapshot $\mathcal{S}_i$ in a sequence of snapshots $\mathcal{G}$ , the model learns representations for the newly introduced entities $\Delta \mathcal{E}_i$ and relations $\Delta \mathcal{R}_i$. These new embeddings are then balanced with the old entity and relation embeddings($\mathcal{E}_{i-1}$ and $\mathcal{R}_{i-1}$), to produce the final representations $\mathcal{E}_{i}$ and $\mathcal{R}_{i}$.

\subsection{Framework}
\begin{figure}[t!]
	\centering
	\includegraphics[width=0.95\linewidth]{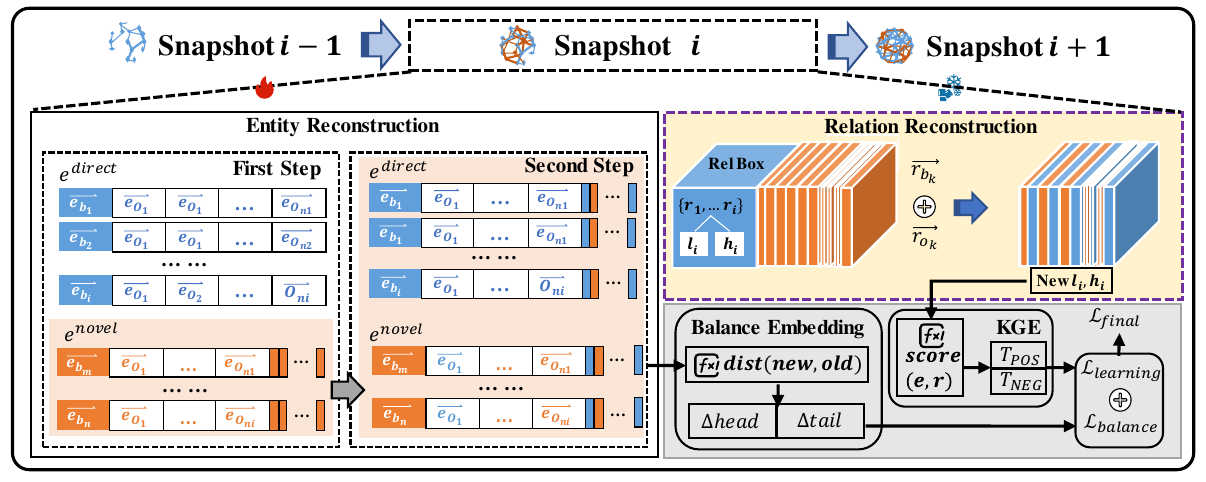} 
	\caption{The Overview of \replaced{STCKGE}{SoTCKGE} framework.}
	\label{fig:2}
\end{figure}
The framework of STCKGE is depicted in Figure \ref{fig:2}. Influenced by previous studies\cite{abboud2020boxe,Gutierrez-Basulto18}, we deviate from the traditional approach of low-dimensional vector space embedding for KGs. Instead, we adopt a spatial-translation-based method. At each phase of KG expansion, we learn embeddings for new knowledge and update the embeddings of existing knowledge using offset vectors. Additionally, the degree of interaction between new and existing knowledge is controlled by a balancing parameters carried over from the prior phase.

\subsection{Embeddings Based on Spatial Translation}
The embedding process consists of two distinct stages. The first stage involves generating the initial knowledge graph embedding for the first snapshot $\mathcal{S}_0$. The second stage iteratively processes subsequent snapshots \replaced{$\{\mathcal{S}_1, \mathcal{S}_2, \ldots ,\mathcal{S}_n\}$}{$\{\mathcal{S}_1, \mathcal{S}_2, \ldots\}$}, embedding newly introduced entities $\Delta\mathcal{E}_i$ while both newly added and existing entities undergo coordinated updates through our bidirectional collaborative update strategy. 
\subsubsection{Initialization Entity And Relation Embeddings}
The knowledge graph embedding for the initial snapshot $\mathcal{S}_0$ \replaced{consists of two components: entity embeddings and relation embeddings.}{comprises entity embeddings and relation embeddings.}  Entity embeddings map each entity to a latent semantic vector space, while relation embeddings model geometric transformations between connected entities. This dual representation captures both structural and relational characteristics, providing a solid foundation for subsequent knowledge integration.

\paragraph{\textbf{Entity Embeddings}}  
Inspired by \cite{abboud2020boxe}, we \replaced{move beyond traditional knowledge graph (KG) embedding methods that represent entities using single low-dimensional vectors. Instead, we adopt a continual entity representation learning framework based on a spatial translation mechanism. This approach helps preserve both the inherent semantics of entities and the dynamic variations caused by their relational contexts.}{abandon the traditional KG embedding method based on low-dimensional vector space and adopt an continual entity representation learning method based on spatial translation mechanism. This method addresses the oversimplification of compressing entities into single low-dimensional vectors. Instead, it uses a dual-component representation framework for entity embedding of initial snapshot entities.} Each entity $\boldsymbol{e}_i$ in the initial snapshot $\mathcal{S}_0$ is represented by \replaced{two components}{a dual-component embedding}: \added{a} base vector $\boldsymbol{e}_i^b \in \mathbb{R}^d$ and \added{an} offset vector $\boldsymbol{e}_i^o \in \mathbb{R}^d$. \replaced{The base vector encodes the static and intrinsic semantics of the entity, while the offset vector captures the influence of its relational neighborhood.}{The base vector $\boldsymbol{e}_i^b$ is a static representation designed to capture the entity's inherent semantics, which remain consistent across different relational contexts. In contrast, the offset vector $\boldsymbol{e}_i^o$ is introduced to dynamically encode contextual variations induced by neighboring relations and entities, enabling relation-aware entity representation.} \added{The final embedding of an entity is defined as the sum of the two components:}
\begin{equation}
\boldsymbol{e}_i = \boldsymbol{e}_i^b + \boldsymbol{e}_i^o
\label{equ:entity_embedding}
\end{equation}

To \replaced{model the local structure of the knowledge graph}{accurately model the topological features of entities in the knowledge graph}, we \replaced{construct}{further construct} a relation-aware entity association set $\Omega_{\boldsymbol{e}_i}$ for each entity\deleted{,} \replaced{. This set contains}{dynamically aggregating} neighboring entities connected through \replaced{various}{different} relation paths. \added{By aggregating information from $\Omega_{\boldsymbol{e}_i}$, the offset vector is updated to reflect contextual changes.} \replaced{This dual-vector representation allows entity embeddings to adapt over time, supporting dynamic knowledge representation and improving the flexibility of the model. The process of generating the final entity representation follows the formulation in Equation~(\ref{equ:1}).}{follows the principle of multi-source information fusion, , overcomes the rigidity of static embeddings and allowing entity representations to dynamically adjust according to their positions in the knowledge graph.}

\begin{equation}
  \label{equ:1}
  \tilde{\boldsymbol{e}_i}  = \boldsymbol{e}_i^b + \sum_{j \in \Omega_{e_i}}\boldsymbol{e}_j^o
\end{equation}

\added{It is worth noting that for isolated entities with $\Omega_{\boldsymbol{e}_i}=\varnothing $, the basis vectors are initialized orthogonally to preserve semantic independence. The offset vectors are initialized with small random noise sampled from a uniform distribution to enable future interaction adaptation.}

\paragraph{\textbf{Relation Embeddings}} 
In addition to embedding entities, we also need to embed the relationships within the knowledge graph. For each relationship $\tilde{\boldsymbol{r}_k}$, we initialize two learnable tensors, $\boldsymbol{r}_k^b$ and $\boldsymbol{r}_k^o$, to define the spatial region represented by the relationship. 
Specifically, for the relationship $\boldsymbol{r}_k \in \mathcal{R}_0$ in the current snapshot $\mathcal{S}_0$ and its corresponding fact triple $(\boldsymbol{e}_i, \boldsymbol{r}_k, \boldsymbol{e}_j) \in \mathcal{T} _i$, where $\boldsymbol{e}_i$ is the head entity, $\boldsymbol{r}_k$ is the relationship, and $\boldsymbol{e}_j$ is the tail entity, we aim to map the head entity $\boldsymbol{e}_i$ to the corresponding tail space through the relationship $\boldsymbol{r}_k$. This mapping process can be implemented as follows:
\begin{enumerate}
  \item \textbf{Initialize the relationship's base point and offset vectors:} For each relationship $r$, we initialize two learnable tensors: the base center point $\boldsymbol{r}_k^b$ and the offset $\boldsymbol{r}_k^o$. The tensor $\boldsymbol{r}_k^b$ represents the reference position of the relationship in the space, while $\boldsymbol{r}_k^o$ indicates the extent of the relationship in the space, starting from the base center point. This offset determines the size of the spatial region represented by the relationship.
  \item \textbf{Calculate the boundaries of the relationship's spatial region:} Using the base center point $\boldsymbol{r}_k^b$ and the offset $\boldsymbol{r}_k^o$, we can calculate the upper and lower boundaries of the relationship's spatial region, denoted as $l_r$ and $u_r$, respectively. The specific formula for this calculation is provided in Equation (\ref{equ:2}) and (\ref{equ:3}).

\begin{align}
  l_r= \min(\boldsymbol{r}_k^b \pm 0.5 \times \boldsymbol{r}_k^o) \label{equ:2}\\u_r= \max(\boldsymbol{r}_k^b \pm 0.5 \times \boldsymbol{r}_k^o)\label{equ:3}
\end{align}

\added{The coefficient 0.5 is inspired by the relational box embedding framework in BoxE\cite{abboud2020boxe}, which applies half-width scaling to define region boundaries. This setting allows a balanced expansion from the center while maintaining geometric interpretability.}
\end{enumerate}
Through these steps, we construct a well-defined spatial region for each relationship in the knowledge graph, enabling more accurate capture and representation of the relationship information.

\subsubsection{Bidirectional Collaborative Update Strategy}
During continual learning for subsequent snapshots, the embedding initialization for newly introduced entities $\Delta \mathcal{E}_i$ at timestamp $i$ follows differentiated \replaced{bidirectional}{hierarchical} rules. New entities are classified into two topological categories: multi-hop topologically independent entities $\boldsymbol{e}_M$ that are entirely independent of the existing knowledge graph (i.e., entities with no direct connections to the outermost layer of the current graph $\mathcal{S}_{i-1}$), and single-hop entities $\boldsymbol{e}_S$ with direct neighborhood relations. 

\paragraph{\textbf{Multi-hop Entity}} 
\replaced{For newly introduced entities that are topologically isolated from the existing knowledge graph}{For newly introduced entities exhibiting topological isolation from the existing knowledge graph } (i.e., structurally disconnected entities), we employ a decoupled representation framework for initialization. This framework establishes a dual-component representation architecture for each newly introduced entity: 1) a base positioning vector $\boldsymbol{e}_M^b$  that encoding core semantic features through orthogonal initialization, and 2) an offset vector $\boldsymbol{e}_M^o$ that dynamically modeling potential interaction patterns with undiscovered facts. \replaced{Meanwhile}{Concurrently}, each new entity is associated with a dedicated relation-entity interaction set $\Omega_{\boldsymbol{e}_i}$ (initialized as $\emptyset$), establishing structural scaffolding for subsequent relational integration. This decoupled initialization design preserves semantic independence through base vectors while enabling adaptive knowledge assimilation \replaced{through}{via} trainable offset components.

\paragraph{\textbf{Single-hop Entity}} 
\replaced{When new entities establish direct connections with the existing knowledge graph, their initialization is guided by a bidirectional neighborhood update mechanism.}{When entities establish direct adjacency relationships with the existing knowledge graph, the initialization process transitions into a bidirectional collaborative update mechanism.} For single-hop entities, we first initialize their base vector $\boldsymbol{e}_S^b$, offset vector $\boldsymbol{e}_S^o$, and relation-entity interaction set  $\Omega_{\boldsymbol{e}_S}$ following the same protocol as for multi-hop entities. Subsequently, a local neighborhood co-update process is activated through two phases:
\subparagraph{\textbf{Old} $\Longrightarrow$ \textbf{New}}
The entity interaction set $\Omega_{\boldsymbol{e}_S}$ undergoes structural augmentation through direct adjacency relationships, incorporating all directly connected entities from $\mathcal{S}_{i-1}$. Concurrently, neighborhood interaction signals dynamically refine the new entity's offset vector via Equation \ref{equ:4}.
 \begin{equation}
  \label{equ:4}
  \tilde{\boldsymbol{e}_S}  \gets \boldsymbol{e}_S^b + \mathbf{BCU}(\boldsymbol{e}_S^o,\{\boldsymbol{e}_{\boldsymbol{e}_j}^o|\boldsymbol{e}_j \in \Omega_{\boldsymbol{e}_S}\}),\quad \boldsymbol{e}_S \in \Delta \mathcal{E}_i
\end{equation}
\subparagraph{\textbf{New} $\Longrightarrow$ \textbf{Old}}
  To maintain the dynamic consistency of the knowledge graph, when the association set of a new entity is updated, a backward update must be performed on the embedding representations of all historical entities within its associated domain. Thanks to the dual-track collaborative representation framework we propose, it is sufficient to apply a low-complexity linear transformation to the offset vectors of the associated old entities through Equation \ref{equ:5}. \replaced{This design allows}{enabling} dynamic adaptation of the embedding space with constant time complexity, \replaced{eliminating }{without} the need \replaced{to}{for} retraining the global representation.
\begin{equation}
  \label{equ:5}
  \tilde{\boldsymbol{e}_j}  \gets \boldsymbol{e}_j^b + \mathbf{BCU}(\boldsymbol{e}_j^o,\boldsymbol{e}_S^o),\quad \boldsymbol{e}_j \in \Omega_{\boldsymbol{e}_S}
\end{equation}

$\mathbf{BCU}(\cdot,\cdot)$ denotes the update strategy we propose for mutual representation refinement, as formalized in Algorithm \ref{alg:bcu}.

\begin{algorithm}[t]
  \caption{Bidirectional Collaborative Update (BCU) }\label{alg:bcu}
  \begin{algorithmic}[1]
      \State Initialize entity $e_S$ with embedding: 
      \State \quad $\emb{e}_S = \emb{e}_S^b + \emb{e}_S^o$ 
      \State Initialize adjacency set: $\setOm \gets \emptyset$ 
      
      \Procedure{\textbf{BCU}}{$\emb{e}_S, \Delta \mathcal{E}_{i-1}$}
          \ForAll{triples $(h, r, t) \in \Delta \mathcal{E}_{i-1}$} 
              \Comment{Iterate new snapshot triples}
              \If{$h = \emb{e}_S.\id$}  
                  \State $\setOm \gets \setOm \cup \{t\}$ 
                  \Comment{Add tail to adjacency set}
              \ElsIf{$t = \emb{e}_S.\id$}
                  \State $\setOm \gets \setOm \cup \{h\}$ 
                  \Comment{Add head to adjacency set}
              \EndIf
          \EndFor
          
          \State Compute updated embedding:
          \State \quad $\widetilde{\emb{e}}_S \gets \emb{e}_S^b + \emb{e}_S^o + \sum_{\emb{e}_j \in \setOm} \emb{e}_j^o$ 
          \Comment{Aggregate neighborhood offsets}
          
          \ForAll{$\emb{e}_j \in \setOm$} 
              \State Update neighbor embedding: 
              \State \quad $\widetilde{\emb{e}}_j \gets \emb{e}_j^b + \emb{e}_j^o + \emb{e}_S^o$ 
              \Comment{Backward propagation}
          \EndFor
          
          \State \Return $\{\widetilde{\emb{e}}_S\} \cup \{\widetilde{\emb{e}}_j\}$ 
          \Comment{Return updated embeddings}
      \EndProcedure
\end{algorithmic}
\end{algorithm}

\subsection{Balancing New and Old Knowledge Embeddings}
\replaced{To enhance the consistency between historical and newly introduced knowledge, we incorporate an additional embedding regularization strategy. This strategy refines the parameter update process by enforcing stricter constraints on the integration of new and old information. Specifically, we calculate the distance between the head and tail entities of both new and historical triples with respect to their associated relations. The model is then optimized to minimize the discrepancy between these distances, thereby promoting stable and balanced representation learning across sequential knowledge snapshots.}{After introducing an additional balanced embedding rule, we refined the parameter update process within the framework to impose stricter constraints and achieve a better balance between the integration of new and old knowledge. To accomplish this, we computed the distance differences between the head and tail entities of new and old facts, relative to their corresponding relationships, and set minimizing these differences as the objective.}

Let a new fact be $(\tilde{\boldsymbol{e}}_{head}^{new}, \boldsymbol{r}^{new},\tilde{\boldsymbol{e}})$, and an old fact be $(\tilde{\boldsymbol{e}}_{head}^{old}, \boldsymbol{r}^{old}, \tilde{\boldsymbol{e}}_{tail}^{old})$. To quantify the embedding differences between new and old knowledge, we define the following metrics:  

\begin{align}
    &\Delta_{head}=  dist(\tilde{\boldsymbol{e}}_{head}^{new},\boldsymbol{r}^{new}) - dist(\tilde{\boldsymbol{e}}_{head}^{old}, \boldsymbol{r}^{old})\\ 
    &\Delta_{tail} = dist(\tilde{\boldsymbol{e}}_{tail}^{new},\boldsymbol{r}^{new}) - dist(\tilde{\boldsymbol{e}}_{tail}^{old}, \boldsymbol{r}^{old})
\end{align}

where $dist(\cdot, \cdot)$ represents a distance computation function that measures the distance between two elements (entities or relationships). This function is specifically designed to account for the associative strength and positional relationships between entities and relationships, as defined below:  

\begin{equation}
  dist(\tilde{\boldsymbol{e}},r)=\frac{|\tilde{\boldsymbol{e}}-c_i|}{w_i^{\delta(e, r)}}-\delta(e,r)\times \varepsilon, \quad \varepsilon=0.5\times(w_i-1)^2\times\frac{w_i+1}{w_i}
\end{equation}

Formally, $\delta(\cdot,\cdot)$ is defined as an indicator function to determine whether a given entity resides within the spatial region demarcated by relation $r$. Specifically, $\delta(\cdot,\cdot)$ is assigned a value of $1$ if and only if the entity satisfies the spatial constraints of relation $r$ (i.e., lies within the spatial boundaries defined by relation $r$); otherwise, it is assigned $-1$. The parameter $w_i$ denotes the width of the relational region boundary, while $c_i$ represents the geometric centroid of the relational space, calculated as follows:
\begin{align}
  c_i=(u_i+l_i)/2\\
  w_i = u_i-l_i+1
\end{align}

Our primary goal is to minimize the defined differences, i.e., to solve $\min(\Delta_{\text{head}}, \Delta_{\text{tail}})$. To comprehensively capture the differences between head and tail entities in the embeddings of new and old knowledge, we further define an overall difference loss function:  

\begin{equation}
    L_{\text{balance}} = \alpha \cdot \Delta_{\text{head}} + \beta \cdot \Delta_{\text{tail}},
\end{equation}

where $\alpha$ and $\beta$ are weighting coefficients used to balance the contributions of the differences in head and tail entities to the loss function. By optimizing this loss function, we aim to minimize the embedding differences between new and old knowledge, thereby achieving a better balance in updating embeddings.

\subsubsection{Overall Objective for Link Predicting}
In the link prediction task, inspired by the core principles of many spatial translation-based embedding methods, we compute scores based on the geometric positions of entity embeddings relative to relation boundaries. This approach avoids the over-reliance on translation-based methods for closely associated facts or those with similar knowledge structures. Specifically, we introduce an innovative loss function, $L_{\text{learn}}$, designed to enhance the model's learning efficiency for distinguishing between positive and negative samples. The calculation is as follows:

\begin{equation}
  \label{equ:10}
   \mathcal{L} _{\text{learn}} = -\log\big(sigmod(\gamma - U_T)\big)- \Delta U \cdot \log\big(sigmod(U_{\bar{T}}  - \gamma)\big)
\end{equation}

where $\Delta U$ serves as the bias weight to balance learning between positive and negative samples, as defined in Equation \ref{equ:11}. The overall evaluation metrics for positive and negative samples, mathematically represented by $U_T$ and $U_{\bar{T}}$ respectively, are formally expressed through the formulations presented in Equations \ref{equ:11}, while $\gamma$ is the margin parameter that regulates the scoring threshold.

\begin{equation}
  \Delta U = U(t)/\sum_{(t )\in T} U(t), \quad U_T = \sum_{t \in T_{\text{pos}}} U(t), \quad U_{\overline{T}} = \sum_{t \in T_{\text{neg}}} U(t)\label{equ:11}
\end{equation}

where $U(t)$ represents the combined score of the head and tail entities for a given fact $t$. 
The composite scoring metric $U(t)$, which quantifies the combined association strength of the corresponding head and tail entities for a given fact $t$, is formulated as follows:

\begin{equation}
U(t)= \exp\big(dist(\text{head}, r) + dist(\text{tail}, r)\big)
\end{equation}

The overall objective function is defined as:
\begin{equation}
\mathcal{L}  = \mathcal{L} _{\text{balance}} + \mathcal{L} _{\text{learn}}.
\end{equation}

\section{Experiments}
\subsection{Datasets}
\replaced{We selected eight datasets for our experiments, ENTITY, RELATION, FACT, HYBRID, GraphEqual, GraphHigher, GraphLower and MULTI. Including seven public datasets: ENTITY, RELATION, FACT, HYBRID are entity-centric, relation-centric, fact-centric, and hybrid\cite{CuiWSLJHH23}; GraphEqual, GraphHigher, GraphLower the increments of triples become equal higher and lower\cite{LiuK0SGLJL24}. Table \ref{tbl:1} and Table \ref{tbl:2} provides detailed statistics for all datasets. }{We selected five datasets for our experiments: ENTITY, RELATION, FACT, HYBRID, and MULTI. The first four are publicly available CKGE datasets, while the latter is a newly constructed dataset. In the construction of the new snapshots for ENTITY, RELATION, FACT, and HYBRID, we primarily added new knowledge related to existing entities. Table \ref{tbl:1} and Table \ref{tbl:2} provides detailed statistics for all datasets.} Each dataset contains five time snapshots, with the training, validation, and test sets split in 3:1:1. 

However, in real-world scenarios, new knowledge may also emerge that is not directly linked to the current knowledge graph, such as multi-hop knowledge. To address this, we extended \replaced{a new multi-hop}{the ENTITY} dataset. Table \ref{tbl:2} summarizes the number of multi-hop and single-hop facts at different snapshots in the MULTI dataset.
\begin{table}[h]
  \centering
  \caption{Datasets statistics.}
  \begin{subtable}[h]{\textwidth}
    \caption{Statistics of four public datasets across different snapshots, ENTITY, RELATION, FACT, and HYBRID. $\Delta E_i, \Delta R_i$ and $\Delta T_i$ denote the sets of cumulative entities, relations and facts in the $i$-th snapshot, respectively.}\label{tbl:1}
    \renewcommand{\arraystretch}{1.2}
    \adjustbox{width=\textwidth}{
      \begin{tabular}{lrrrrrrrrrrrrrrr}
      \toprule
      \multirow{2}{*}{Dataset} & \multicolumn{3}{c}{Snapshot 0}&\multicolumn{3}{c}{Snapshot 1}&\multicolumn{3}{c}{Snapshot 2}&\multicolumn{3}{c}{Snapshot 3}&\multicolumn{3}{c}{Snapshot 4} \\ \cmidrule(lr){2-4} \cmidrule(lr){5-7} \cmidrule(lr){8-10} \cmidrule(lr){11-13} \cmidrule(l){14-16}  &$\Delta E_0$ & $\Delta R_0$ & $\Delta T_0$&$\Delta E_1$ & $\Delta R_1$ & $\Delta T_1$ &$\Delta E_2$ & $\Delta R_2$ & $\Delta T_2$&$\Delta E_3$ & $\Delta R_3$ & $\Delta T_3$&$\Delta E_4$&$\Delta R_4$ & $\Delta T_4$ \\ 
      \midrule
      ENTITY     & 2,909  & 233  & 46,388 & 5,817  & 236   & 72,111 & 8,275  & 236 & 73,785  & 11,633 & 237 & 70,506  & 14,541 & 237 & 47,326 \\
      RELATION  & 11,560 & 48   & 98,819 & 13,343 & 96    & 93,535 & 13,754 & 143 & 66,136  & 14,387 & 190 & 30,032  & 14,541 & 237 & 21,594 \\
      FACT       & 10,513 & 237  & 62,024 & 12,779 & 237   & 62,023 & 13,586 & 237 & 62,023  & 13,894 & 237 & 62,023  & 14,541 & 237 & 62,023 \\
      HYBRID     & 8,628  & 86   & 57,561 & 10,040 & 102   & 20,873 & 12,779 & 151 & 88,017  & 14,393 & 209 & 103,339 & 14,541 & 237 & 40,326 \\
      GraphEqual &2,908   &226   &57,636  &5,816   &235    &62,023  &8,724   &237   &62,023   &11,632 &237  &62,023   &14,541  &237  & 66,411\\
      GraphHigher &900    &197   &10,000  &1,838   &221    &20,000  &3,714   &234   &40,000   &7,467  &237  &80,000   &14,541  &237  &160,116 \\
      GraphLower &7,505   &237   &160,000 &11,258  &237    &80,000  &13,134  &237   &40,000   &14,072  &237 &20,000   &14,541  &237  &10,116\\
      \bottomrule
      \end{tabular}
      }
  \end{subtable}
  \begin{subtable}[h]{\textwidth}
    \vspace*{0.5em}
    \centering
    \caption{MULTI statistics. Since Snapshot 4 represents the whole knowledge graph, the number of multi-hop facts is zero.}\label{tbl:2}
    \adjustbox{width=0.5\textwidth}{
    \begin{tabular}{lrrrrr}
    \toprule
    \multirow{2}{*}{MULTI} & \multirow{2}{*}{Entity} & \multirow{2}{*}{Relation} & \multicolumn{3}{c}{Fact} \\
    \cmidrule(lr){4-6}
    & & & total & single-hop & multi-hop \\
    \midrule
    Snapshot 0 & 3504 & 188 & 38324 & 37946 & 378 \\
    Snapshot 1 & 6516 & 226 & 35090 & 34743 & 347 \\
    Snapshot 2 & 9707 & 237 & 82199 & 81386 & 813 \\
    Snapshot 3 & 12385 & 237 & 82082 & 81270 & 812 \\
    Snapshot 4 & 14541 & 237 & 72421 & 72421 & 0 \\
    \bottomrule
    \end{tabular}
    }
  \end{subtable}  
\end{table}


\subsection{Experiment Settings}

\subsubsection{Baselines.}
\replaced{We carefully selected 11 baseline models from five perspectives, including one non-continual learning model, two models based on dynamic architectures, three memory replay-based methods, three regularization-based models and two distillation-based training models.}{We carefully selected 11 baseline models for comparison, including one non-continual learning method, Finetune \cite{CuiWSLJHH23}; two models based on dynamic architectures, PNN \cite{rusu2016progressive} and CWR \cite{lomonaco2017core50}; three memory replay-based methods, GEM \cite{lopez2017gradient}, EMR \cite{DBLP:conf/naacl/WangXYGCW19}, and DiCGRL \cite{KouLLLZZ20}; three regularization-based models, SI \cite{zenke2017continual}, EWC \cite{kirkpatrick2017overcoming}, and LKGE \cite{CuiWSLJHH23}; and two distillation-based training methods, FastKGE \cite{LiuK0WGSL0JL24} and IncDE \cite{LiuK0SGLJL24}.}

\begin{itemize}
  \item \textbf{Non-continual learning:} We include Fine-tune \cite{CuiWSLJHH23}, which simply fine-tunes the old model on new data without countering catastrophic forgetting.
  \item \textbf{Dynamic-architecture methods:} We consider PNN \cite{rusu2016progressive} and CWR \cite{lomonaco2017core50}; these approaches expand the model only for newly added entities or relations.
  \item \textbf{Memory-replay methods:} We select GEM \cite{lopez2017gradient}, EMR \cite{DBLP:conf/naacl/WangXYGCW19}, and DiCGRL \cite{KouLLLZZ20}; they mitigate forgetting by replaying stored samples or learned prototypes.
  \item \textbf{Regularization-based methods:} We adopt SI \cite{zenke2017continual}, EWC \cite{kirkpatrick2017overcoming}, and LKGE \cite{CuiWSLJHH23}; these techniques constrain parameter drift with additional loss terms.
  \item \textbf{Distillation-based methods:} We use FastKGE \cite{LiuK0WGSL0JL24} and IncDE \cite{LiuK0SGLJL24}, which retain past knowledge through various knowledge-distillation objectives.
\end{itemize}
\subsubsection{Evaluation Metrics.}
To systematically evaluate the model's performance, we conducted rigorous experimentation on the link prediction task through the following protocol: (1) For each factual triple in the test set, we systematically replaced either the head entity or tail entity through uniform random sampling from the entity space; (2) Generated candidate entity sequences by enumerating all possible substitutions while maintaining the original relation constraints; (3) Computed ranking scores using the learned embeddings and sorted candidates accordingly.

To ensure methodological rigor, we implemented seven complementary evaluation metrics:
\begin{itemize}
  \item \textbf{Accuracy Metrics:} Employed Mean Reciprocal Rank (MRR) and Hit@$k$ ($k\in\{1,10\}$) to quantify ranking precision.
  \item \textbf{Knowledge Transfer Capacity:} Assessed through Forward Transfer Time (FWT) and Backward Transfer Time (BWT) \cite{CuiWSLJHH23}, measuring temporal knowledge integration efficiency.
  \item \textbf{Efficiency:} Recorded end-to-end inference latency across the entire test set.
\end{itemize}

\subsubsection{Implementation details.}
All experiments were implemented using PyTorch on a NVIDIA RTX 4060Ti GPU. The time parameter $i$ was selected from the range [0, 1, 2, 3, 4], and Adam was used as the optimizer. The learning rate was set to $1\times 10^{-4}$, chosen from the candidate set [$1\times 10^{-4}, 5\times 10^{-4}, 8\times 10^{-4}$] while the batch size was dynamically adjusted within the range [512, 1024, 2048]. To ensure fairness and reliability, all experimental results were reported as the average of five runs. Moreover, for all control experiments, excluding those involving our proposed model, TransE \cite{TransE} was used as the baseline KGE model.

\begin{table}[h]
  \centering
  \caption{Experimental results obtained on the test sets across all snapshots of ENTITY, RELATION, FACT, HYBRID,GraphEqual, GraphHigher, GraphLower and MULTI. The bolded results represent the best performance.}\label{tbl:3}
  \renewcommand{\arraystretch}{1.2}
  \begin{minipage}[h]{\textwidth}
  \adjustbox{width=\textwidth}{
  \begin{tabular}{l|ccc|ccc|ccc|ccc}
  \hline
  & \multicolumn{3}{c|}{\textbf{ENTITY}} & \multicolumn{3}{c|}{\textbf{RELATION}} & \multicolumn{3}{c|}{\textbf{FACT}} & \multicolumn{3}{c}{\textbf{HYBRID}} \\
  \textbf{Method} & MRR & Hits@1 & Hits@10 & MRR & Hits@1 & Hits@10 & MRR & Hits@1 & Hits@10 & MRR & Hits@1 & Hits@10 \\
  \toprule
  Fine-tune\cite{CuiWSLJHH23} & 0.165 & 0.085 & 0.321 & 0.093 & 0.039 & 0.195 & 0.172 & 0.090 & 0.339 & 0.135 & 0.069 & 0.262 \\

  PNN\cite{rusu2016progressive}& 0.229 & 0.130 & 0.425 & 0.167 & 0.096 & 0.305 & 0.157 & 0.084 & 0.029 & 0.185 & 0.101 & 0.349 \\
  CWR\cite{lomonaco2017core50} & 0.088 & 0.028 & 0.202 & 0.021 & 0.10 & 0.043 & 0.083 & 0.030 & 0.192 & 0.037 & 0.015 & 0.077 \\

  GEM\cite{lopez2017gradient} & 0.165 & 0.085 & 0.321 & 0.093 & 0.040 & 0.196 & 0.175 & 0.092 & 0.345 & 0.136 & 0.070 & 0.263 \\
  EMR\cite{DBLP:conf/naacl/WangXYGCW19} & 0.171 & 0.090 & 0.330 & 0.111 & 0.105217 & 0.225 & 0.171 & 0.090 & 0.337 & 0.141 & 0.073 & 0.267 \\
  DiCGRL\cite{KouLLLZZ20} & 0.107 & 0.057 & 0.211 & 0.133 & 0.079 & 0.241 & 0.162 & 0.084 & 0.320 & 0.149 & 0.083 & 0.277\\

  SI\cite{zenke2017continual} & 0.154 & 0.072 & 0.311 & 0.113 & 0.055 & 0.224 & 0.172 & 0.088 & 0.343 & 0.111 & 0.049 & 0.229 \\
  EWC\cite{kirkpatrick2017overcoming} & 0.229 & 0.130 & 0.423 & 0.165 & 0.093 & 0.306 & 0.201 & 0.113 & 0.382 & 0.186 & 0.102 & 0.350 \\
  LKGE\cite{CuiWSLJHH23} & 0.234 & 0.136 & 0.425 & 0.192 & 0.106 & 0.366 & 0.210 & 0.122 & 0.387 & 0.207 & 0.121 & 0.379 \\

  FastKGE\cite{LiuK0WGSL0JL24} & 0.239 & 0.146 & 0.427 & 0.185 & 0.107 & 0.359 & 0.203 & 0.117 & 0.384 & 0.211 & 0.128 & 0.382 \\
  IncDE\cite{LiuK0SGLJL24} & 0.253 & 0.156 & \textbf{0.448} & 0.199 & 0.106 & 0.370 & 0.216 & 0.131 & \textbf{0.391} & 0.224 & 0.131 & \textbf{0.401} \\
  \midrule
  STCKGE & \textbf{0.269} & \textbf{0.206} & 0.398 & \textbf{0.244} & \textbf{0.180} & \textbf{0.374} & \textbf{0.249} & \textbf{0.192} & 0.367 & \textbf{0.248} & \textbf{0.189} & 0.361 \\
  \bottomrule
  \end{tabular}
  }
  \end{minipage}
  \begin{minipage}[h]{\textwidth}
  \vspace*{0.5em}
  \centering
  \renewcommand{\arraystretch}{1.2}
  \adjustbox{width=\textwidth}{
  \begin{tabular}{l|ccc|ccc|ccc|ccc}
  \hline
  & \multicolumn{3}{c|}{\textbf{GraphEqual}} & \multicolumn{3}{c|}{\textbf{GraphHigher}} & \multicolumn{3}{c|}{\textbf{GraphLower}} & \multicolumn{3}{c}{\textbf{MULTI}} \\
  \textbf{Method} & MRR & Hits@1 & Hits@10 & MRR & Hits@1 & Hits@10 & MRR & Hits@1 & Hits@10 & MRR & Hits@1 & Hits@10 \\
  \toprule
  Fine-tune\cite{CuiWSLJHH23} & 0.183 & 0.096 & 0.358 & 0.198 & 0.108 & 0.375 & 0.185 & 0.098 & 0.363 & 0.202 & 0.117 & 0.375 \\

  PNN\cite{rusu2016progressive} & 0.212 & 0.118 & 0.405 & 0.186 & 0.097 & 0.364 & 0.213 & 0.119 & 0.407 & 0.149 & 0.086 & 0.273 \\
  CWR\cite{lomonaco2017core50} & 0.122 & 0.041 & 0.277 & 0.189 & 0.096 & 0.374 & 0.032 & 0.005 & 0.080 & 0.180 & 0.098 & 0.345 \\

  GEM\cite{lopez2017gradient} & 0.189 & 0.099 & 0.372 & 0.197 & 0.109 & 0.372 & 0.170 & 0.084 & 0.346 & 0.201 & 0.116 & 0.372 \\
  EMR\cite{DBLP:conf/naacl/WangXYGCW19} & 0.185 & 0.099 & 0.359 & 0.202 & 0.113 & 0.379 & 0.188 & 0.101 & 0.362 & 0.183 & 0.105 & 0.342 \\
  DiCGRL\cite{KouLLLZZ20} & 0.104 & 0.040 & 0.226 & 0.116 & 0.041 & 0.242 & 0.102 & 0.039 & 0.222 & 0.174 & 0.104 & 0.370 \\

  SI\cite{zenke2017continual} & 0.179 & 0.092 & 0.353 & 0.190 & 0.099 & 0.371 & 0.186 & 0.099 & 0.366 & 0.194 & 0.107 & 0.370 \\
  EWC\cite{kirkpatrick2017overcoming} & 0.207 & 0.113 & 0.400 & 0.202 & 0.106 & 0.385 & 0.210 & 0.116 & 0.405 & 0.182 & 0.102 & 0.347 \\
  LKGE\cite{CuiWSLJHH23} & 0.214 & 0.118 & 0.407 & 0.207 & 0.120 & 0.382 & 0.210 & 0.116 & 0.403 & 0.215 & 0.126 & 0.373 \\

  FastKGE\cite{LiuK0WGSL0JL24} & 0.229 & 0.124 & 0.370 & 0.222 & 0.127 & 0.403 & 0.218 & 0.124 & 0.405 & 0.154 & 0.102 & 0.248 \\
  IncDE\cite{LiuK0SGLJL24} & 0.234 & 0.134 & 0.432 & 0.227 & 0.132 & 0.412 & 0.228 & 0.129 & \textbf{0.426} & 0.195 & 0.110 & 0.357 \\
  \midrule
  STCKGE & \textbf{0.288} & \textbf{0.215} & \textbf{0.436} & \textbf{0.331} & \textbf{0.253} & \textbf{0.489} & \textbf{0.251} & \textbf{0.186} & 0.385 & \textbf{0.324} & \textbf{0.249} & \textbf{0.476} \\
  \bottomrule
  \end{tabular}
  }
  \end{minipage}  
\end{table}

\subsection{Results}
\subsubsection{Main Results}
\deleted{Table \ref{tbl:3} presents experimental results on eight datasets. Our model demonstrates strong competitive performance across most of datasets. Furthermore, the model outperforms all baseline models on the multi-hop dataset, indicating its superior ability to learn multi-hop relationships. It also shows strong competitive performance on the four publicly available single-hop datasets.}

\deleted{First, STCKGE demonstrated outstanding performance on the multi-hop dataset MULTI and the publicly available dataset FACT, while also exhibiting strong competitive performance on three additional datasets. Notably, on the MULTI, STCKGE achieved significant improvements in MRR, Hits@1, Hits@3, and Hits@10, with increases of 5.7\%, 6.7\%, 6.7\%, and 7.9\%, respectively. This exceptional performance strongly supports the robustness and effectiveness of using relations as region embeddings for multi-hop relationship learning, even in continual learning scenarios.}

\deleted{Moreover, on the publicly available single-hop dataset FACT, STCKGE showed performance improvements over the best method, with gains of 0.1\%, 4.7\%, 0.2\%, and 1.5\% in MRR, Hits@1, Hits@3, and Hits@10, respectively. Particularly noteworthy is the fact that STCKGE outperformed all baseline models in the Hits@1 metric, with improvements ranging from 2.1\% to 6.7\%, further highlighting its higher accuracy and reliability.}

\deleted{Overall, STCKGE demonstrated average improvements of 0.14\%, 4.44\%, 1.94\%, and 1.68\% in MRR, Hits@1, Hits@3, and Hits@10, respectively, across five datasets compared to the best baseline. While STCKGE maintained consistent growth in Hits@1 and Hits@3 on the three additional public datasets, with improvements ranging from 2.1\% to 6.1\% and 0.2\% to 2.0\%, respectively, the MRR metric showed a decline on the ENTITY and HYBRID datasets. We hypothesize that this decline may be due to the baseline models using TransE as the underlying knowledge graph embedding model, which primarily models relationships based on vector distances between entities. In contrast, STCKGE incorporates relation-based region embeddings, which allow for more complex interactions. When faced with ambiguous or polysemous relationships, this difference may cause STCKGE to focus more on optimizing the most probable answer positions, rather than uniformly improving the ranking of all relevant answers.}

\added{To evaluate STCKGE in continual knowledge graph embedding (CKGE), we conduct comprehensive experiments on eight datasets. STCKGE outperforms all eleven representative baselines across every main metric, particularly when the graph structure evolves rapidly. The results are summarized in Table \ref{tbl:3}.}

\added{Overall, STCKGE maintains a stable lead. Across eight datasets and 24 evaluation metrics, it achieves significant improvements on 20 metrics. STCKGE shows notable advantages in MRR and Hits@1 on most datasets. On GraphEqual, GraphHigher, and GraphLower, where the rate of new triples is deliberately balanced, increased, or decreased, STCKGE improves MRR by 2.3-10.4\% and Hits@1 by 5.7-2.1\%. In the more challenging ENTITY, RELATION, FACT, and HYBRID settings, STCKGE achieves an average gain of 3.0\% in MRR and 6.1\% in Hits@1, with the largest improvements reaching 4.5\% and 7.4\%, respectively. On the multi-hop dataset MULTI, STCKGE surpasses all baselines across every metric. These consistent and substantial gains across diverse tasks and metrics demonstrate the strong generalizability of the proposed approach.}

\added{The greater the structural evolution, the more substantial STCKGE's advantage. In ENTITY and RELATION, where entities and relations are frequently added, traditional replay- or regularization-based methods (EWC, SI, GEM) struggle to expand the embedding space and fall behind incremental distillation approaches (IncDE, FastKGE). STCKGE effectively preserves discriminative information from historical structures while offering an expandable basis for new entities and relations, significantly mitigating catastrophic forgetting. In FACT and HYBRID, where facts are frequently added or removed but the overall structure remains relatively stable, replay methods remain competitive. However, STCKGE still outperforms them, with FACT-MRR increasing from 0.216 to 0.249 and HYBRID-MRR improving from 0.224 to 0.248. Finally, on GraphEqual, GraphHigher, GraphLower, and MULTI, STCKGE improves Hits@1 over the strongest baseline by 5.4\%, 10.4\%, 2.3\%, and 12.9\%, respectively, demonstrating robustness to diverse evolution patterns.}

\subsubsection{Ablation Results}
To quantitatively assess the efficacy of our proposed \replaced{Bidirectional Collaborative Update}{Hierarchical Embedding Update} strategy and the Balanced Embedding method, we constructed two ablated variants:
\begin{itemize}
  \item  \textbf{w/o \replaced{BCU}{HEU}:} Disabling the \replaced{bidirectional collaborative Update}{hierarchical embedding update} strategy.
  \item \textbf{w/o BE:} Disabling the balanced embedding method.
\end{itemize}

As empirically demonstrated in Table \ref{tbl:4}, the ablation of either component induces statistically significant performance degradation. The BE mechanism exhibits particularly pronounced impacts, manifesting average relative reductions of 3.9\% (MRR), 2.96\% (Hits@1), 4.74\% (Hits@3), and 3.58\% (Hits@10) across the five benchmark datasets, comprising four established single-hop datasets (ENTITY, RELATION, FACT, HYBRID) and our new multi-hop dataset (MULTI). This empirical evidence substantiates two critical insights:
(1)The staged embedding propagation mechanism plays a pivotal role in facilitating gradient-controlled knowledge integration. (2)The equilibrium regularization, while exhibiting relatively smaller effect magnitudes (average $\Delta$ MRR=1.2\%), effectively prevents representation collapse through eigenvalue stabilization.


\begin{table}[t]
  \caption{Ablation results on ENTITY, RELATION, FACT, HYBRID and MULTI.}
  \label{tbl:4}
  \renewcommand{\arraystretch}{0.5}
  \begin{minipage}[t]{\textwidth}
      \centering
      \adjustbox{width=\textwidth}{
      \begin{tabular}{lcccccccccccccc}
      \toprule
      \multirow{2}{*}{Method} & \multicolumn{4}{c}{ENTITY}&\multicolumn{4}{c}{RELATION}&\multicolumn{4}{c}{FACT} \\ \cmidrule(lr){2-5} \cmidrule(lr){6-9} \cmidrule(lr){10-13} & MRR & Hits@1 &Hits@3 &Hits@10 & MRR & Hits@1 &Hits@3 &Hits@10 & MRR & Hits@1 &Hits@3 &Hits@10  \\ 
      \midrule
      STCKGE & 0.198 & 0.172 & 0.276 & 0.405     & 0.203 & 0.166 & 0.234 & 0.366     & 0.217 & 0.172 & 0.269 & 0.393 \\
      w/o \replaced{BCU}{HEU} & 0.194 & 0.163 & 0.247 & 0.381     & 0.1925 & 0.1438 & 0.181 & 0.3544  & 0.1802 & 0.1521 & 0.247 & 0.384\\
      w/o BE & 0.157 & 0.152 & 0.239 & 0.372      & 0.1712 & 0.123 & 0.179 & 0.3353   & 0.1529 & 0.1632 & 0.239 & 0.372 \\
      \bottomrule         
    \end{tabular}
    }
\end{minipage}
\begin{minipage}[t]{\textwidth}
  \vspace*{0.5em}
  \centering
  \adjustbox{width=0.7\textwidth}{
  \begin{tabular}{lcccccccccccccc}
    \toprule
    Method & \multicolumn{4}{c}{HYBRID} & \multicolumn{4}{c}{MULTI} \\
    \cmidrule(lr){2-5} \cmidrule(lr){6-9}
    & MRR & Hits@1 & Hits@3 & Hits@10 & MRR & Hits@1 & Hits@3 & Hits@10 \\
    \midrule
    STCKGE  & 0.211 & 0.156 & 0.254 & 0.414    & 0.212 & 0.172 & 0.276 & 0.429 \\
    w/o HEU  & 0.184 & 0.151 & 0.207 & 0.391    & 0.177 & 0.163 & 0.247 & 0.392\\
    w/o BE   & 0.121 & 0.1302 & 0.176 & 0.378   & 0.162 & 0.152 & 0.239 & 0.371\\
    \bottomrule
  \end{tabular}
  }
\end{minipage}
\end{table}

\subsection{\added{Method Analyze}}
\subsubsection{\added{Computational Complexity Analysis.}}
\added{Let $|\Delta E_i|$ denote the number of newly introduced entities and $|\Delta T_i|$ the number of new triples in the $i$-th snapshot. The average node degree is denoted by $k$ (typically, $k \ll |E|$), and $d$ is the embedding dimension. During the initialization phase, a one-time linear projection is applied to the $|E_0| + |R_0|$ entity and relation vectors in the initial snapshot $\mathcal{S}_0$, resulting in a time complexity of $\mathcal{O}((|E_0| + |R_0|) \cdot d)$. For each subsequent snapshot, the HEU procedure traverses the adjacency list. Since each new entity is associated with at most $k$ existing entities, both forward and backward updates require $\mathcal{O}(|\Delta E_i| \cdot k \cdot d)$ time. Embedding alignment involves computing the distance difference between new and historical triples. When mini-batch sampling is used with batch size $b$, the computation cost is reduced to $\mathcal{O}(b \cdot d)$. Overall, the worst-case time complexity per snapshot is $\mathcal{O}(|T| \cdot d)$. However, in real-world sparse graphs, this degrades to $\mathcal{O}((|\Delta E_i| + |\Delta T_i|) \cdot k \cdot d)$, which matches the time complexity of translation-based continual KGE methods. As for space complexity, the model only needs to store $(|E| + |R|) \cdot d$ embedding vectors and $|T|$ triples, leading to a total space complexity of $\mathcal{O}((|E| + |R|) \cdot d + |T|)$.}
\subsubsection{MULTI Prediction Capability}
\begin{figure}[h]
	\centering
  \includegraphics[width=0.75\textwidth]{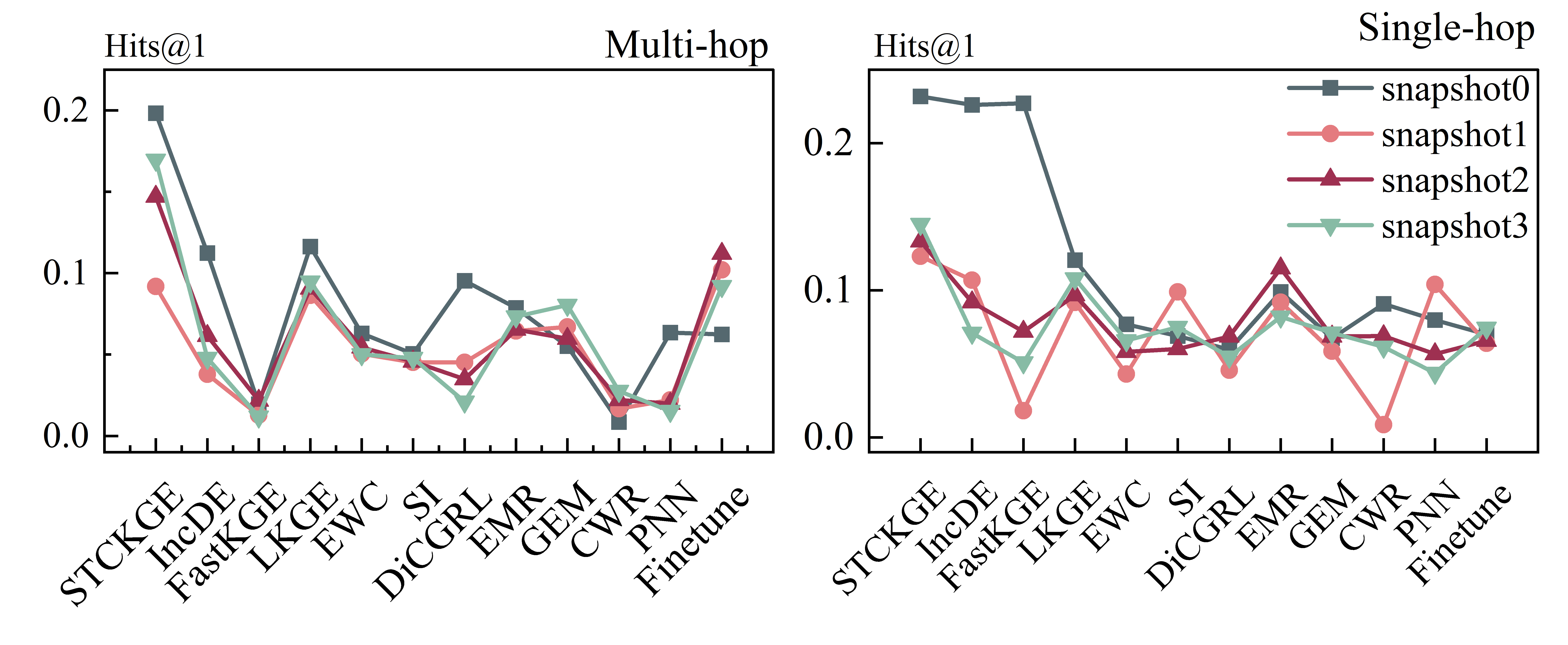}
	\caption{ Results represent the probability distribution for multi-hop and single-hop relationships, when model achieves the Hits@1 accuracy on the test sets of the four snapshots in the MULTI dataset.}
	\label{fig:3}
\end{figure}

As shown, on the MULTI dataset, STCKGE outperforms all baseline models across four evaluation metrics. To assess the model's ability to predict multi-hop relationships, we considered the characteristics of the dataset construction. In the first four snapshots, the ratio of single-hop to multi-hop relationships is relatively balanced, so the model trained on the fifth snapshot was evaluated using the Hits@1 score on the test sets of the first four snapshots. The number of facts in the test sets for the first four snapshots are 7666(67), 7018(60), 16441(164), and 16417(151), with the numbers in parentheses representing the multi-hop relationships. The results, shown in Figure \ref{fig:3}, demonstrate that STCKGE exhibits stronger capabilities in handling multi-hop facts compared to the other baseline models.

\subsubsection{Learning Performance Over Time}
\begin{figure}[h]
	\centering
	\includegraphics[width=0.85\textwidth]{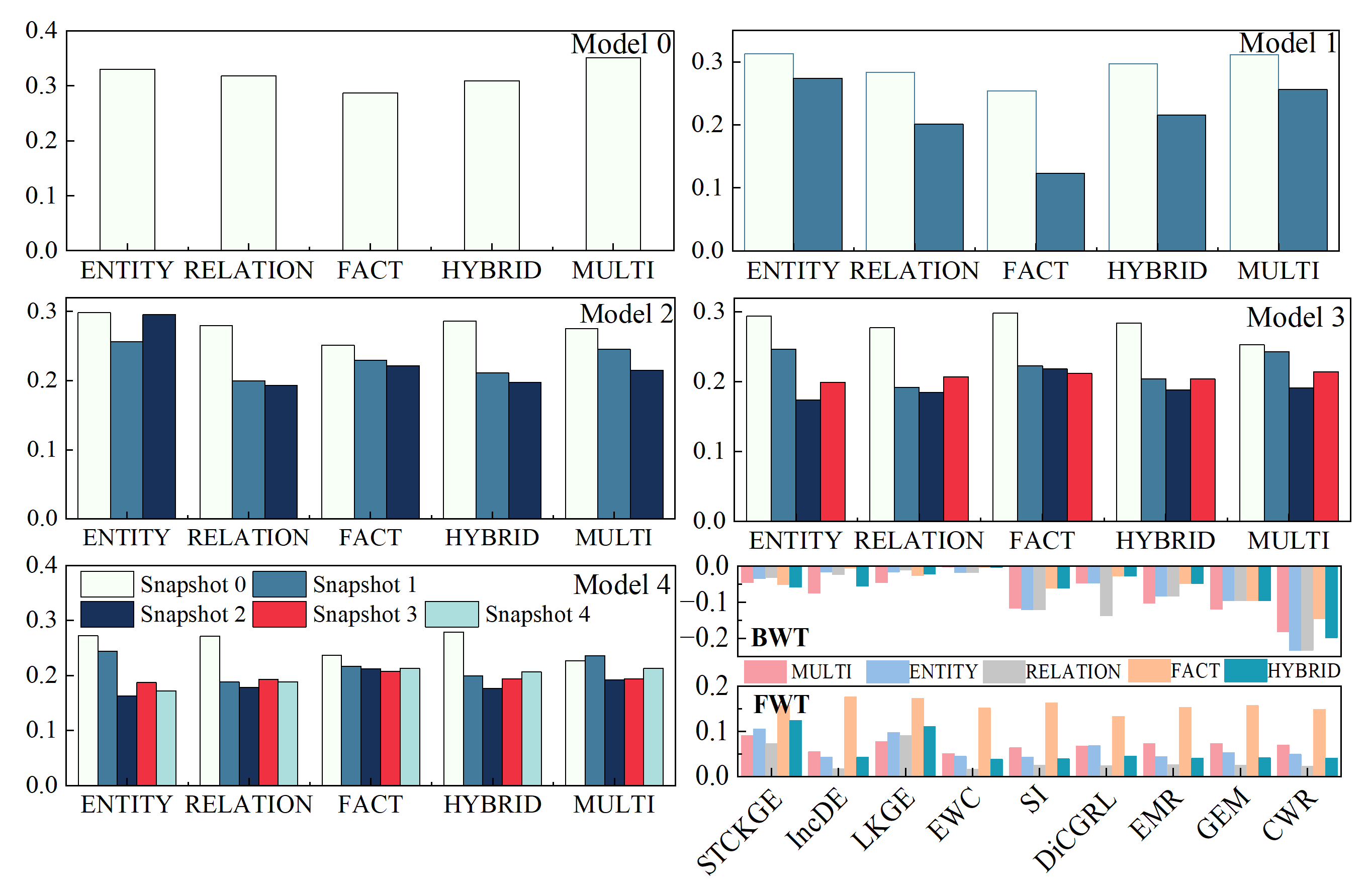} %
	\caption{Results of STCKGE's dynamic learning capability at each time. Model $i$ represents the model trained on the $i$-th snapshot.}
	\label{fig:4}
\end{figure}

The last figure in Figure \ref{fig:4} illustrates the dynamic learning capability of our model over time. Specifically, in Model 3 and Model 4, as time progresses and knowledge accumulates, the model snapshots at different time points are able to maintain stable learning performance. This phenomenon clearly demonstrates the ability of STCKGE to balance the integration of new and old knowledge, effectively absorbing new information while preserving existing knowledge, thus ensuring a smooth fusion and embedding of both.The comparison results of FWT and BWT across different models and datasets demonstrate that, on the multi-hop dataset MULTI, STCKGE outperforms all baseline models in knowledge transfer ability. Moreover, it remains highly competitive on the other four public datasets.

\subsubsection{Effectiveness of Embedding Update Strategy.}

\begin{figure}[h]
  \centering 
  \begin{subfigure}[h]{0.48\textwidth} 
    \centering
  \includegraphics[width=0.8\linewidth]{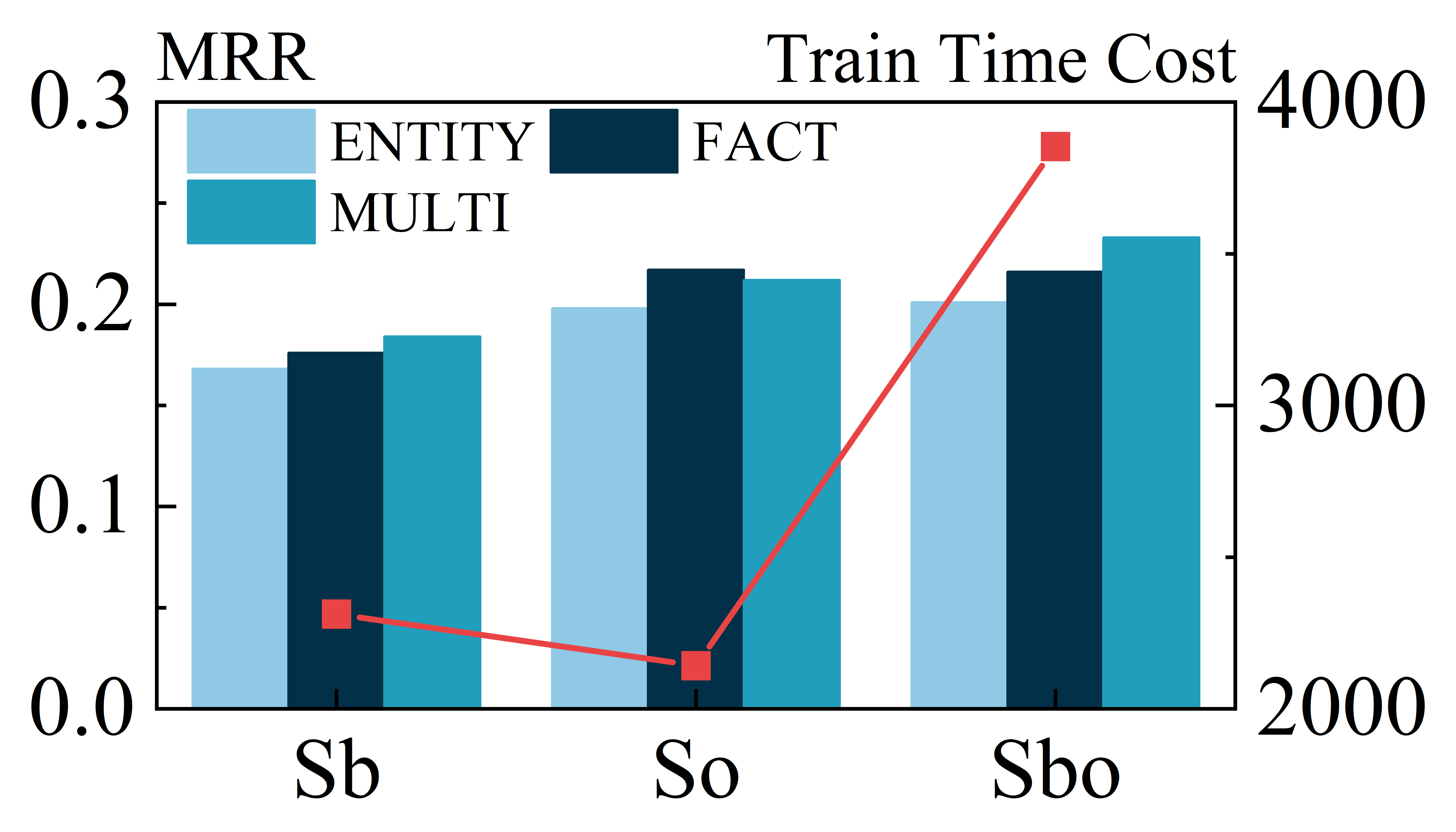}
    \caption{Effectiveness of different methods for embedding update.}
    \label{fig:6}
  \end{subfigure}
  \begin{subfigure}[h]{0.48\textwidth} 
    \centering
    
    \includegraphics[width=0.85\linewidth]{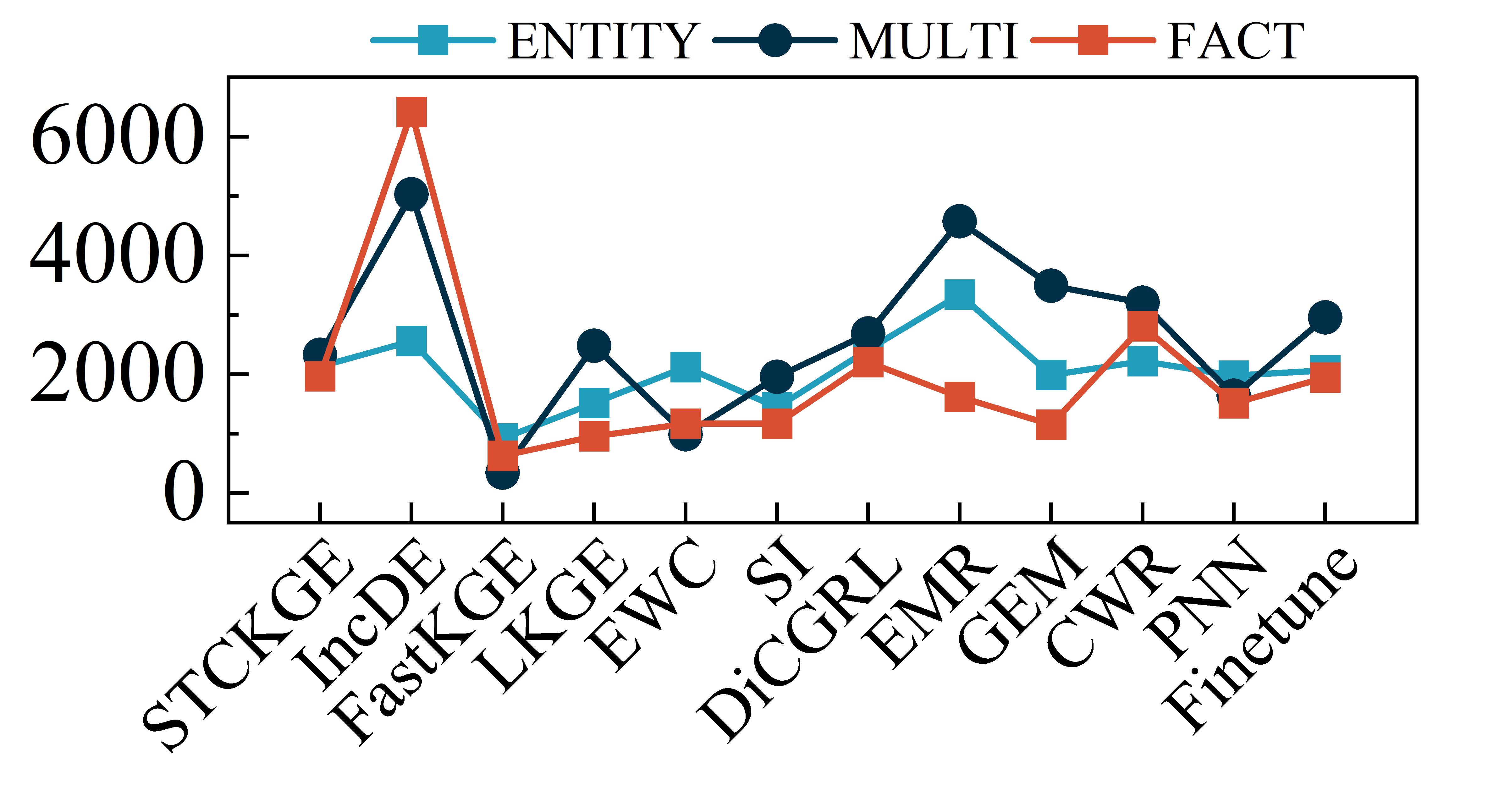} 
    \caption{Train time cost on the ENTITY, FACT, and MULTI datasets.}
    \label{fig:7}
  \end{subfigure}
  \caption{Results on Effectiveness of Embedding Update Strategy and Learning Efficiency.}
\end{figure}

We conducted experiments on three datasets to evaluate the impact of different embedding update methods on model performance, as shown in the Figure \ref{fig:6}. $Sb$ denotes the update of entity base vectors only, $So$ represents the update of offset vectors only, and $Sbo$ indicates the simultaneous update of both entity base vectors and offset vectors. As shown, the $Sbo$ method yields better performance, but it incurs significant time overhead. To better balance model performance and efficiency, we ultimately chose to update only the offset vectors for the new and old entities.
    
\subsubsection{Learning Efficiency}
To comprehensively evaluate the efficiency of the STCKGE model, we compared its training time overhead with all baseline models across the ENTITY, FACT, and MULTI datasets, as shown in Figure \ref{fig:7}. The results indicate that the training time of STCKGE on these datasets is comparable to that of other baseline models using continual  learning or fine-tuning methods. Notably, when processing the more complex MULTI dataset, the training time of most models increases significantly, whereas the increase in STCKGE's training time is relatively modest. This phenomenon highlights the advantage of the STCKGE model—it can handle more complex tasks in real-world scenarios while maintaining efficient performance.
\subsubsection{Limitation}
\added{Although STCKGE posts strong MRR and Hits@1 scores, its Hits@10 trails the best baseline by 3.8\% in averageon ENTITY, FACT, GraphLower and HYBRID, suggesting that the region-based geometry still struggles to surface enough high confidence candidates for tail queries; moreover, the gap widens when snapshots are sparse, indicating that the balancing loss may over-regularise low-frequency entities and that further architectural tweaks, such as learnable relation-region annealing or adaptive margin scheduling could unlock additional gains without sacrificing efficiency.}
\section{Conclusion}

In this paper, we propose a new CKGE framework based on relation-based regional embeddings. In this framework, the position of an entity is jointly controlled by a base position vector and an offset vector. This design not only enhances the model's ability to represent complex relational structures, but also enables efficient embedding updates of both new and old knowledge through simple spatial transformations, without relying on continual  learning methods. Additionally, we introduce a \replaced{bidirectional collaborative Update}{hierarchical embedding update} strategy and a balanced embedding method to refine the parameter update process. This approach not only reduces training costs but also improves model accuracy. Future research will further explore the integration of the relation-based regional CKGE framework with continual  learning methods to enhance both learning efficiency and accuracy.

\section*{Acknowledgments}

This work was supported by the China National Key R\&D Plan [No. 202208060371 and No. 2023YFB3106900] and the National Science Foundation of China [No. 62372334].





\bibliographystyle{elsarticle-num} 
\bibliography{refs}






\end{document}